# Measurement of collisions between laser cooled cesium atoms and trapped cesium ions

Sourav Dutta[1,2,][*] and S. A. Rangwala[2]

[1]*Tata Institute of Fundamental Research, 1 Homi Bhabha Road, Colaba, Mumbai 400005, India*
[2]*Raman Research Institute, C. V. Raman Avenue, Sadashivanagar, Bangalore 560080, India*

We report the measurement of collision rate coefficient for collisions between ultracold Cs atoms and low energy $Cs^+$ ions. The experiments are performed in a hybrid trap consisting of a magneto-optical trap (MOT) for Cs atoms and a Paul trap for $Cs^+$ ions. The ion-atom collisions impart kinetic energy to the ultracold Cs atoms resulting in their escape from the shallow MOT and, therefore, in a reduction in the number of Cs atoms in the MOT. By monitoring, using fluorescence measurements, the Cs atom number and the MOT loading dynamics and then fitting the data to a rate equation model, the ion-atom collision rate is derived. The Cs-$Cs^+$ collision rate coefficient $9.3(\pm0.4)(\pm1.2)(\pm3.5) \times 10^{-14}$ m³s⁻¹, measured for an ion distribution with most probable collision energy of 95 meV ($\approx k_B.1100\,K$), is in fair agreement with theoretical calculations. As an intermediate step, we also determine the photoionization cross section of Cs $6P_{3/2}$ atoms at 473 nm wavelength to be $2.28(\pm0.33) \times 10^{-21}$ m².

## I. INTRODUCTION

The measurement of ion-atom collisions has been pursued for many years but was mostly restricted to high collision energies ($E > 1$ eV) [1–3] until the advent of hybrid traps, where atoms and ions are trapped simultaneously [4–13] [14,15]. In these hybrid traps, the ions are typically trapped in a Paul trap while the cold atoms are trapped either in a magneto-optical trap (MOT) or in an optical dipole trap (ODT). Typically, the atoms are at a temperature of around 0.1 mK while the ions are at a higher temperature ranging from around 1 mK to a few 1000 K depending on whether or not the ions are cooled. These hybrid traps have enabled the study of atom-ion collisions at energies well below 1 eV ($\equiv k_B.11605\,K$). In addition, the population and the quantum state of atoms and ions in these traps can also be experimentally controlled, making these systems quite versatile and suitable for studies of charge exchange processes [5,8,12], cold chemical reactions [16,17], ion cooling mechanisms [6,10,18], non-equilibrium processes [13,19,20] and charge transport [21,22]. In this article we use a hybrid trap for the measurement of low energy Cs-$Cs^+$ collisions. Notably, this system is interesting because Cs has the largest polarizability among alkali-metal atoms, making the ion-atom collision cross section for Cs-$Cs^+$ the largest.

The ion-atom interaction potential at long range is given by $V = -C_4/2R^4$, where $R$ is the inter-nuclear separation and $C_4$ is a constant that is directly proportional to the polarizability ($\alpha$) of the atom. The ion-atom collision cross section $\sigma$ ($\propto C_4^{2/3}E^{-1/3}$) [22] depends on the value of $C_4$ and therefore a measurement of $\sigma$ provides a measure of $C_4$. Note that $\sigma$ also depends on the collision energy $E$ which in turn is proportional to $v^2$, where $v$ is relative ion-atom speed. The advent of hybrid traps allows studies in the low $E$ regime, which is difficult in ion beam experiments. In order to determine the ion-atom collision cross section, we measure the ion-atom collision rate $\gamma_{ia}$, which is essentially the product of collision cross-section $\sigma$ and the ion flux $n_i v$ (the number of ions crossing per unit area per unit time). Here $n_i$ is the density of trapped ions which we estimate from experiments in conjunction with computer simulations. Under certain conditions mentioned later in the text, the collision rate is expressed as $\gamma_{ia} = n_i \langle \sigma v \rangle$. The angular bracket denotes a weighted average over all speeds accessed in the experiment. With $\gamma_{ia}$ and $n_i$ determined, we can derive $\langle \sigma v \rangle$ which is called the collision rate coefficient and is denoted by $k_{ia}$. The experimentally measured $k_{ia}$ can then be compared with theoretically calculated $k_{ia}$.

The methodology used for measurement of the ion-atom collision rate [23–25] bears semblance with the method used to measure atom-atom collision rates in MOTs [26,27]. However, in this article we use a different experimental protocol compared to Ref. [23] so that the standard phenomenological rate equation model used to derive the collision rate coefficient is followed. The basic concept of the experiment is the following. The ultracold Cs atoms, at a temperature below 1 mK, are trapped in a MOT, which typically has a trap depth of the order of a few Kelvin i.e. few 100 μeV. The Cs atoms in the MOT are predominantly in the $6S_{1/2}$, $F = 4$ state, with negligible population in $6S_{1/2}$, $F = 3$ state and a small but significant fraction in the $6P_{3/2}$, $F' = 5$ state. The $Cs^+$ ions are trapped in a much deeper Paul trap with depth of ~ 0.8 eV and ions are relatively much hotter (~ 1100 K). Thus, the $Cs^+$ ions have large speeds while the ultracold Cs atoms in the MOT are essentially at rest – an ion-atom collision therefore invariably transfers kinetic energy to the atom. For overwhelmingly large fraction of the elastic and resonant charge exchange collisions [22], the energy transferred to the atom in a single collision is much higher than the MOT depth, resulting in the atom escaping from the MOT after a collision. This leads to a change in the steady-state atom number as well as the loading dynamics of the MOT, which can be used to determine the atom-ion collision rate $\gamma_{ia}$. We monitor the atom number using fluorescence measurement and record the loading dynamics of the Cs MOT in the presence and absence of trapped Cs$^+$ ions and then fit the data to a phenomenological rate equation model to determine $\gamma_{ia}$. We then determine $k_{ia}$ using the expression $k_{ia} = \gamma_{ia}/n_i$ and compare it with theoretical calculations. As an intermediate step, we also determine the photoionization cross section of Cs $6P_{3/2}$ atoms at 473 nm wavelength and compare with previous reports [28–33].

*sourav.dutta@tifr.res.in

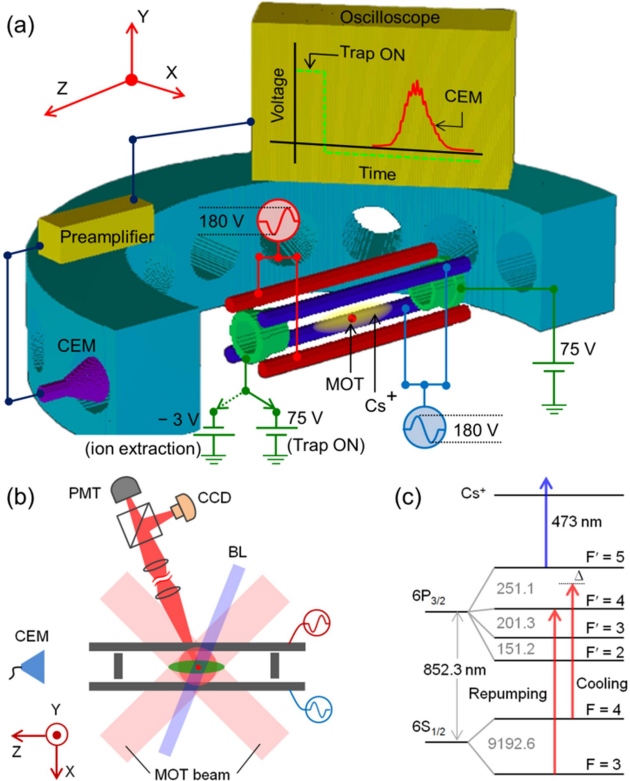

FIG. 1. (a) A schematic of the experimental setup. The Paul trap consists of four rod electrodes biased with rf voltages and two annular end cap electrodes biased with dc voltages. The centers of the Paul trap and the MOT are well overlapped. The size of the MOT is much smaller than the ion cloud. The CEM in combination with a preamplifier and an oscilloscope is used to detect ions extracted from the ion trap. (b) Schematic top view of the setup showing the directions of the MOT beams (two pairs in the x-z plane and one pair in the y direction), the fluorescence detection system and the blue laser (BL) beam. (c) The energy levels of Cs atoms relevant to the experiment. The numbers in grey are the frequency difference (in MHz) between consecutive $F$ levels. The cooling, repumping and ionizing lasers are indicated.

Note that *purely* atomic hyperfine state changing collisions, i.e. collisions that change the hyperfine state of the Cs atom from the $F = 4$ to the $F = 3$ state while imparting negligible kinetic energy to the atoms, do not lead to appreciable loss of atoms because Cs atoms in the $F = 3$ state are rapidly optically pumped back to the $F = 4$ state where they are trapped. However, we can legitimately ignore contributions from such inelastic collisions that impart negligible kinetic energy to the atoms while changing, for example, the Cs atomic hyperfine state because the likelihood of such collisions is extremely small.

The manuscript is organized as follows. In the next section we provide a brief description of the hybrid atom-ion trap. In section III we present the experimental measurements and discuss the phenomenological rate equation model used to determine the ion-atom collision rate. We conclude in section IV with a discussion.

## II. EXPERIMENTAL SETUP

The apparatus consists of a MOT for ultracold Cs atoms and a Paul trap for Cs$^+$ ions such that the centers of the two traps overlap (see Fig. 1). The Cs MOT requires two lasers near 852.3 nm – a cooling laser, whose frequency is -16.5 MHz detuned from the cycling $6S_{1/2}$, $F = 4 \leftrightarrow 6P_{3/2}$, $F' = 5$, and a repumping laser, whose frequency is on resonance with the $6S_{1/2}$, $F = 3 \leftrightarrow 6P_{3/2}$, $F' = 4$ transition. The cooling and repumping lights are combined using a beam splitter such that all MOT beams have light at both frequencies. The Cs MOT is formed by three pairs of mutually orthogonal retro-reflected laser beams that intersect at the center of the vacuum chamber, where a magnetic field gradient of ~ 15 Gauss/cm exists. The switching on/off of the MOT beams is controlled using mechanical shutters placed in the beam path. The Cs MOT atoms are detected by monitoring their fluorescence. A calibrated photomultiplier tube (PMT) is used to infer the atom number while a CCD camera is used to record the atomic density profile. In absence of ions, the MOT typically has ~ $9 \times 10^6$ atoms at a density of ~ $4.5 \times 10^{10}$ cm$^{-3}$. The $1/e^2$ radius of the MOT ($r_a$) is ~ 0.47 mm. As is usual for MOTs, a fraction of the atoms is in the excited $6P_{3/2}$ state. This fraction ($f_e$) is determined to be 0.092(12) from the expression $f_e = \left(\frac{1}{2}\right) \frac{(I_c/I_{sat})}{1+4(\Delta/\Gamma)^2+(I_c/I_{sat})}$ for an ideal two-level system, where $\Gamma = 2\pi \cdot 5.234$ MHz is the natural linewidth of the $6P_{3/2}$ state, $\Delta = -16.5$ ($\pm 1.0$) MHz is the measured detuning of the cooling laser, $I_c = 25$ mW/cm$^2$ is the intensity of the cooling laser (measured with a power meter of accuracy 3%) and $I_{sat} = 2.71$ mW/cm$^2$ is the saturation intensity (assuming the light to have isotropic polarization in the MOT region) [34].

The Paul trap consists of four rod electrodes and a pair of annular end cap electrodes. Each of the rod electrodes has a diameter of 3 mm and the center-to-center distance between adjacent rods is 13 mm. The design and operation has been discussed in an earlier report on trapped Rb$^+$ ions [9]. In order to trap Cs$^+$ ions, radio-frequency (rf) voltage of amplitude 90 V and frequency 430 kHz is applied to the rod electrodes such that adjacent rods are biased oppositely (see Fig. 1). This provides the radial confinement for the ions. The axial confinement is provided by DC voltage of 75 V applied to the two end cap electrodes. The Paul trap is loaded by photo-ionization of Cs MOT atoms using a blue laser of wavelength $\lambda_p$ = 473 nm which ionizes Cs atoms from the $6P_{3/2}$ state [Fig. 1(c)]. The $1/e^2$ radius of the blue laser beam ($r_{BL}$) is 1.41 mm, significantly larger than the $1/e^2$ radius of the MOT, to provide approximately uniform illumination across the MOT atoms. The collimated blue laser beam is chosen over a light emitting diode (LED) based diverging light source to mitigate the effect of light induced atomic desorption (LIAD) of Cs atoms from the inner walls of the vacuum chamber and the glass viewports. A diverging light source such as a LED could alter the MOT loading rate and atom number, leading to uncertainties in the measurement of $\gamma_{ia}$.

The Cs$^+$ ions do not have transitions in the optical frequency regime and cannot be detected using fluorescence detection. Therefore, the number ($N_i$) of trapped Cs$^+$ ions is measured by extracting the ions from the trap and detecting them on a channel electron multiplier (CEM). The CEM cone voltage is kept at -1800 V. The output of the CEM is fed to a low-pass preamplifier which generates a signal proportional to the number of detected ions. The ion extraction is performed by suddenly switching one of the end cap voltages to -3 V [see Fig. 1(a)]. The ions arrive at the CEM detector after a time of flight (ToF) that depends on the position and velocity of ion just prior to extraction, creating an arrival time distribution that is measured on an oscilloscope [see Fig. 1(a)]. The area under the curve is proportional to the number of ions detected, where the proportionality constant is derived as follows. Keeping the number of trapped ions fixed at a low value, the CEM output is recorded in two modes – one in which the output is fed to a preamplifier and the other in which the output is directly fed to the oscilloscope. In the latter case, each ion detected gives a single pulse of width 8 ns and therefore the number of ions detected can be directly counted (additionally, the CEM cone voltage was kept at higher magnitude -2450 V to increase the signal to noise ratio of the pulses). Comparing this number with the area under the curve of the former case provides the calibration. A similar strategy for calibration was used in Ref. [35] and CEM saturation effects were seen to be negligible for $N_i \sim O(10^5)$ in Ref. [24]. Note that the number of *detected* ions is lower than the number of *trapped* ions due to two reasons. The first reason is that some of the trapped ions may not hit the CEM upon extraction. However, such ion loss is typically small and we assume, based on computer simulations in SIMION®, that all trapped ions make it to the CEM. The second factor is the sub-unity quantum efficiency $\eta$ of the detector, which we assume to be 0.37 ± 0.05 from the specifications of the CEM manufacturer. Therefore, we multiply the number of *detected* ions by $1/\eta$ to arrive at the number of *trapped* ions $N_i$.

## III. EXPERIMENTAL RESULTS AND ANALYSIS

Three separate experimental sequences are involved in the determination of the ion-atom collision rate. The first records the loading of a Cs MOT alone [Fig. 2(a)], the second records the loading of a Cs MOT in the presence of the blue ionization laser [Fig. 2(b)] and the third records the loading of a Cs MOT in the presence of trapped ions [Fig. 2(c)]. The fitted parameters from all three experiments are then combined to extract the ion-atom collision rate. In what follows, we will describe each experiment and the corresponding rate equation, one at a time.

### A. Loading of a Cs MOT

We record the loading dynamics of the Cs MOT as follows. The magnetic field gradient for the MOT is always kept on for all experiments reported here. The Cs cooling and repumping lights are allowed to enter the vacuum chamber at time $t = 0$ by unblocking the mechanical shutter in the beam path. A typical loading curve is shown in Fig. 2(a). The loading dynamics of a MOT can be modelled using a rate equation of the form:

$$\frac{dN_a}{dt} = L_a - \gamma_b N_a - \beta_a \int n_a^2 \, d^3r \qquad (1)$$

where $N_a$ is number of atoms in the MOT, $n_a$ is the number density of the atoms, $L_a$ is the rate at which atoms are loaded into the MOT (depends on the background flux of atoms, laser intensities and detunings etc.), $-\gamma_b N_a$ accounts for loss of MOT atoms due to collisions with a constant-density background gas (depends on the background vapour pressure in the vacuum chamber) and $-\beta_a \int n_a^2 \, d^3r$ accounts for loss of MOT atoms due to two-body collisions among the MOT atoms [26,27]. Our MOT operates in a regime where the volume $V_a$ ($\sim 0.43 \times 10^{-9}$ m$^3$) of the MOT remains approximately constant as $N_a$ changes, allowing us to make the approximation $\beta_a \int n_a^2 \, d^3r \approx (\beta_a/V_a)N_a^2$. Clearly, this term becomes more important as $N_a$ increases. However, previous reports suggest that $\beta_a$ is small ($\sim 10^{-12}$ cm$^3$s$^{-1}$) for our MOT parameters [36], resulting in $(\beta_a/V_a)N_a^2 \sim 0.023 \times 10^7$ s$^{-1}$ for the highest value of $N_a$ ($\sim 10^7$) used in our experiment. This value of $(\beta_a/V_a)N_a^2$ is much smaller than $\gamma_b N_a$ ($\sim 0.24 \times 10^7$ s$^{-1}$, determined below) allowing us to neglect the $\beta_a \int n_a^2 \, d^3r$ term in Eq. (1). The rate equation then becomes:

$$\frac{dN_a}{dt} = L_a - \gamma_b N_a. \quad (2)$$

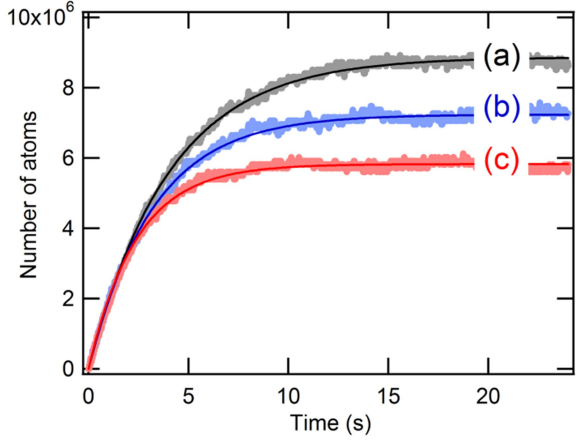
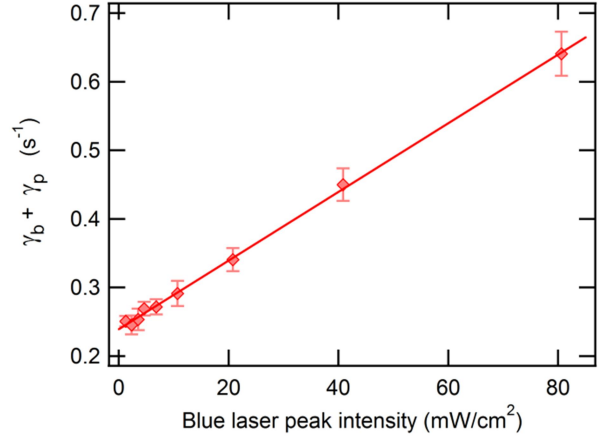

FIG. 2. Typical MOT loading curves along with the fits to the rate equation model. (a) Loading of a Cs MOT in absence of the ionization laser. This data is used to determine the values of $L_a$ and $\gamma_b$. (b) Loading of a Cs MOT in presence of the ionization laser with intensity 10.7 mW/cm$^2$. This data is used to determine the value of $\gamma_p + \gamma_b$, from which $\gamma_p$ is extracted by subtracting $\gamma_b$. (c) Loading of a Cs MOT in presence of the ionization laser and trapped Cs$^+$ ions. This data is used to determine the value of $\gamma_p + \gamma_b + \gamma_{ia}$, from which $\gamma_{ia}$ is extracted by subtracting $\gamma_p + \gamma_b$.

FIG. 3. The values $\gamma_b + \gamma_p$ plotted against the intensity $I_p$ of the blue ionization laser. The slope of the linear fit yields the value of $\zeta (= \gamma_p/I_p)$, while the y-intercept gives the value of $\gamma_b$.

The solution to Eq. (2) is:

$$N_a = \frac{L_a}{\gamma_b}(1 - e^{-\gamma_b t}). \quad (3)$$

A typical loading curve along with a fit to Eq. (3) is shown in Fig. 2(a). We obtain $L_a = 2.20(\pm 0.11) \times 10^6$ atoms/s and $\gamma_b = 0.241(\pm 0.011)$ s$^{-1}$ from the fit. If neglecting the $\beta_a \int n_a^2 \, d^3r$ term was not a valid assumption, one would observe a departure of the experimental data from Eq. (3), especially at large $N_a$ i.e. during the later stages of MOT loading curve. However, no such departure is visible in our experimental data [see Fig. 2(a)] confirming the validity of the assumption.

### B. Ionization of Cs atoms

The Cs$^+$ ions are created by photo-ionization of Cs MOT atoms using a blue laser of wavelength 473 nm. To determine the effect of photo-ionization on the MOT atoms, the MOT loading is recorded in presence of the blue laser while keeping the ion trap voltages off. A typical loading curve is shown in Fig. 2(b). The rate equation in this case is:

$$\frac{dN_a}{dt} = L_a - \gamma_b N_a - \gamma_p N_a \quad (4)$$

where $-\gamma_p N_a$ accounts for loss of MOT atoms due to photo-ionization. The solution to Eq. (4) is:

$$\overline{N}_a = \frac{L_a}{\gamma_b + \gamma_p}\left(1 - e^{-(\gamma_b + \gamma_p)t}\right). \quad (5)$$

We fit the experimental data to Eq. (5) to obtain $L_a$ and $(\gamma_b + \gamma_p)$. A typical fit is shown in Fig. 2(b). The $L_a$ thus obtained is equal to that obtained in section III.A where the blue laser was absent. This implies that the blue laser does not alter the MOT loading rate. We perform the experiment at different intensities ($I_p$) of the blue laser and, for each $I_p$, obtain $(\gamma_b + \gamma_p)$ from a fit to Eq. (5). The plot of $(\gamma_b + \gamma_p)$ vs. $I_p$ is shown in Fig. 3. The linear dependence of $(\gamma_b + \gamma_p)$ arises from the linear dependence of $\gamma_p$ on $I_p$, while the y-intercept yields $\gamma_b$ which is independent of $I_p$. The slope of the curve is $\zeta = \gamma_p/I_p = 5.0(\pm 0.3) \times 10^{-4}$ J$^{-1}$m$^2$.

The value of $\zeta$ can be used to determine the photo-ionization cross section ($\sigma_p$) of Cs atoms in 6$P_{3/2}$ state for a photo-ionization laser wavelength of 473 nm. This determination of $\sigma_p$ is, however, not required for the evaluation of ion-atom collision rate $\gamma_{ia}$ or the ion-atom collision rate coefficient $k_{ia}$. For low values of $I_p$, i.e. far below the saturation photo-ionization intensity, $\gamma_p$ can be expressed as [29]:

$$\gamma_p = f_e \frac{\lambda_p \sigma_p I_p}{hc} = \zeta I_p \quad (6)$$

where $h$ is the Planck's constant, $c$ is the speed of light, $\lambda_p$ = 473 nm is the wavelength of photo-ionization laser and $f_e$ = 0.092(12) is the fraction of Cs MOT atoms in the $6P_{3/2}$ state. Using the measured value of $\zeta$, we get $\sigma_p$ = $2.28(\pm 0.33) \times 10^{-21}$ m$^2$. This value is in the same range as previous experimental measurements [28–30] and theoretical calculations [31–33] which report $\sigma_p$ in the range $1.5 - 2.0 \times 10^{-21}$ m$^2$.

We note that there could be an additional experimental systematic error, arising from a systematic error in the estimate for $f_e$, which if present would lower the value of $\sigma_p$ determined above. The value $f_e = 0.092$ was obtained considering Cs atom to be an ideal two-level system and assuming $I_{sat} = 2.71$ mW/cm$^2$, the saturation intensity for *isotropic polarization* of the cooling light [34]. We expect Cs to behave similar to an ideal two-level system, in comparison with other alkali-metal atoms, since the hyperfine splitting in Cs is larger than that in all other alkali-metal atoms. In addition, given the conditions in a MOT where light polarization direction and magnetic field direction are not constant in space, the assumption of *isotropic polarization* seems valid and is typically used by the community [34,37]. However, as mentioned in [34] "…this is almost certainly an overestimate of the effective saturation intensity, since sub-Doppler cooling mechanisms will lead to optical pumping and localization in the light maxima". For example, the saturation intensity for linearly (circularly) polarized light is 1.66 (1.10) mW/cm$^2$ and would result in $f_e = 0.135$ (0.178) and $\sigma_p = 1.56 \pm 0.09 \times 10^{-21}$ ($1.18 \pm 0.07 \times 10^{-21}$) m$^2$.

We also note that $f_e$ enters in the expression for determination of MOT atom number (and therefore in the determination of the MOT loading rate). However, uncertainties in determination of $f_e$ does not lead to errors in the measurement of $\gamma_b$, $\gamma_p$ or $\gamma_{ia}$ since all the rate equations, i.e. Eqs. (2), (4) and (7), are linear functions of the number of MOT atoms and the common factor $f_e$ cancels out.

### C. Atom-ion collisions

To measure the ion-atom collision rate $\gamma_{ia}$, a protocol similar to the one used in Ref. [24] is followed. The MOT is first loaded to a steady state with the ion trap voltages and the blue laser kept on. This also loads the ion trap with Cs$^+$ ions and the number of trapped ions reaches as steady state within a few seconds. The MOT cooling and repumping beams are then blocked for a brief period of 300 ms during which the Cs MOT empties out completely, but the number of trapped ions $N_i$ and ion density $n_i$ ($\approx N_i/V_i$) remains almost unchanged. Here $V_i$ is volume occupied by the trapped ions and will be assumed to be independent of $N_i$. The cooling and repumping beams are then turned back on and the loading of the Cs MOT in presence of the trapped ions and the blue laser is recorded. The rate equation in this case is:

$$\frac{dN_a}{dt} = L_a - \gamma_b N_a - \gamma_p N_a - k_{ia} \int n_i n_a d^3r. \quad (7)$$

The last term in Eq. (7) accounts for loss of MOT atoms due to collisions with trapped ions and $k_{ia}$ is the ion-atom collision rate coefficient. In writing Eq. (7) we have implicitly assumed that each ion-atom collision results in the loss of the atom from the MOT. This is a valid assumption since the trap depth of MOT is small (~ 0.1 meV) and the kinetic energy imparted to the atom in a collision with an ion, which typically has large kinetic energy (~ 95 meV, see section III.E.), is much higher than the MOT depth.

Since $n_i$ is kept approximately constant while the MOT loads and $V_i \gg V_a$, we can write $k_{ia} \int n_i n_a d^3r \approx k_{ia} n_i \int n_a d^3r = k_{ia} n_i N_a = \gamma_{ia} N_a$, where $\gamma_{ia} = k_{ia} n_i$ is the ion-atom collision rate. With this substitution, the solution to Eq. (7) is:

$$\bar{\bar{N}}_a = \frac{L_a}{\gamma_b + \gamma_p + \gamma_{ia}} \left(1 - e^{-(\gamma_b + \gamma_p + \gamma_{ia})t}\right). \quad (8)$$

A typical MOT loading curve in the presence of trapped ions along with the fit to Eq. (8) is shown in Fig. 2(c). We obtain $L_a$ and $(\gamma_b + \gamma_p + \gamma_{ia})$ from the fit. We again find that $L_a$ thus obtained is equal to that in section III.A, confirming that the presence of the blue laser and trapped ions do not alter $L_a$. Since $\gamma_b$ and $\gamma_p$ are already known from sections III.A and III.B, the value of $\gamma_{ia}$ can be calculated from the value of $(\gamma_b + \gamma_p + \gamma_{ia})$.

We then perform the experiment at different intensities ($I_p$) of the blue laser. Different values of $I_p$ result in different steady state values of $N_i$ and $N_a$, but the MOT atom number obeys the same set of rate equations. We determine $\gamma_{ia}$ for each value of $I_p$ and plot $\gamma_{ia}$ vs. $I_p$ in Fig. 4. The initial increase in $\gamma_{ia}$ with increase in $I_p$ is due to increase in ion number $N_i$ and hence density $n_i$ of the trapped ions with increase in $I_p$. Eventually the value of $\gamma_{ia}$ saturates because of saturation of $n_i$ at large $I_p$. To verify this saturation effect we measured $N_i$ by extracting the ions from the trap and detecting them using the CEM. The plot in Fig. 5 shows that $N_i$ indeed varies with $I_p$ and tends to saturate at large $I_p$, similar to the trend in Fig. 4. The saturation of $N_i$ (and hence $n_i$) can, in principle, occur due to two reasons. The first is that the repulsive ion-ion interaction starts playing a role as $n_i$ increases and therefore limits the number of ions that can be accommodated in the ion trap which has a finite depth. This effect is small because ion-ion interaction is small in our experiment as explained in section III.D. The second reason is that, in a steady state ion trap, the loading rate of ions into the ion trap $L_i$ depends on the number of available atoms $\bar{\bar{N}}_a^\infty$ [obtained by setting $t \to \infty$ in Eq. (8)] and the photo-ionization rate $\gamma_p$, i.e. $L_i \approx \bar{\bar{N}}_a^\infty \gamma_p$. Both $\gamma_p$ and $\bar{\bar{N}}_a^\infty$ depend on $I_p$ and the functional forms are obtained from Eqs. (6) and (8), respectively. Using these in $L_i \approx \bar{\bar{N}}_a^\infty \gamma_p$ and setting $I_p \to \infty$, we find that $L_i \to constant$ i.e. the ion loading rate $L_i$ cannot increase indefinitely. This limits the number of ions and the ion density in the ion trap.

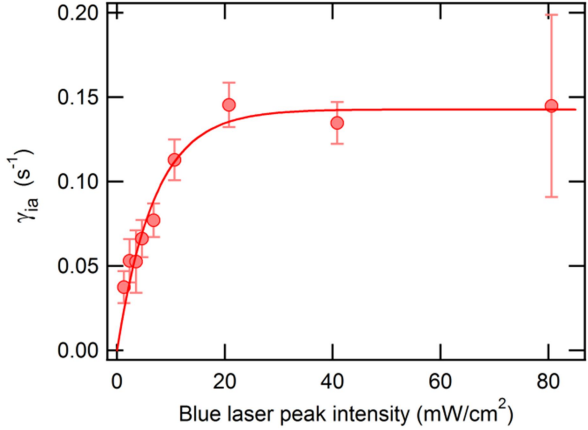

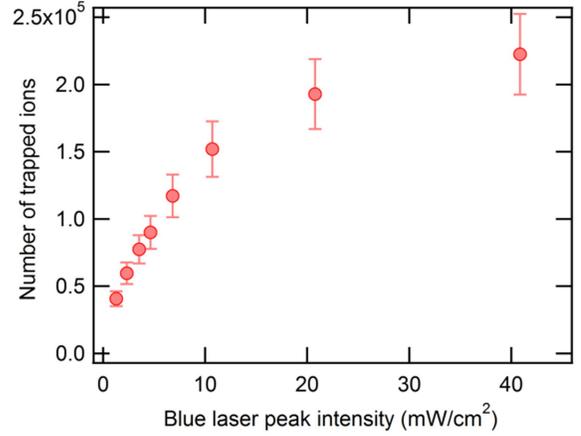

FIG. 4. The atom-ion collision rates $\gamma_{ia}$ at different intensities $I_p$ of the blue ionization laser. Also shown is a fit to $\gamma_{ia} = \gamma_{ia}^{max}(1 - e^{-I_p/I_{p,0}})$, yielding $\gamma_{ia}^{max} = 0.143(6)$ s$^{-1}$ and $I_{p,0} = 7.0(8)$ mW/cm$^2$. The value of $\gamma_{ia}$ initially increases due to increase in the number of trapped ions $N_i$ and then saturates since $N_i$ saturates at large $I_p$. The error bar of the data point at the highest intensity is large because the steady-state values of $\bar{N}_a$ and $\bar{\bar{N}}_a$ are small, leading to relatively large atom number fluctuations.

FIG. 5. The number of trapped ions $N_i$ estimated from the ion signal detected on the CEM after the ions are extracted from the ion trap. Initially $N_i$ increases with $I_p$ and then tends to saturate.

We have justified and validated all approximations and assumptions made so far and the experimental data fits very well with the rate equation model. This gives us the confidence that the determination of the atom-ion collision rate $\gamma_{ia}$ is quite robust, accurate and devoid of *systematic* errors. In what follows, we need to determine the values of $N_i$ and $V_i$, both of which have larger error margins. This increases the error budget in the determination of $k_{ia}$ as explained below.

### D. Atom-ion collision rate coefficient

The atom-ion collision rate coefficient $k_{ia}$ can be derived from the expression $k_{ia} = \gamma_{ia}/n_i \approx (\gamma_{ia}/N_i)V_i$. The value of $\gamma_{ia}/N_i$ can be determined by combining the data in Figs. 4 and 5, and plotting $\gamma_{ia}$ as a function of $N_i$ as shown in Fig. 6. From the slope of the $\gamma_{ia}$ vs. $N_i$ plot, we determine $\gamma_{ia}/N_i = 6.9(\pm 0.3)(\pm 0.9) \times 10^{-7}$ s$^{-1}$, where the value in the first parenthesis represents the 1σ uncertainty band of the fit and the value in the second parenthesis accounts for the 13% uncertainty in $N_i$ arising from errors in the estimation of quantum efficiency $\eta$ of the detector.

The quantity that still remains to be determined is the volume $V_i$ occupied by the trapped ions. Since the Cs$^+$ ions are optically dark, their spatial distribution cannot be imaged using light. Two methods [23,24] have previously been used to determine $V_i$ of such ions – computer simulations of the ion trap and dependence of $\gamma_{ia}/N_i$ on the MOT displacement with respect to the ion trap center. Both methods are prone to systematic errors. In case of simulations, it is hard to ensure that it matches the experiment exactly while in the case of MOT displacement, it is hard to ensure that the MOT atom number, density profile and overlap with the ionization laser remain unchanged during the measurement. In our apparatus, we found it impossible to change the position of the MOT without affecting its shape and size significantly. We will therefore use a computer simulated value for $V_i$.

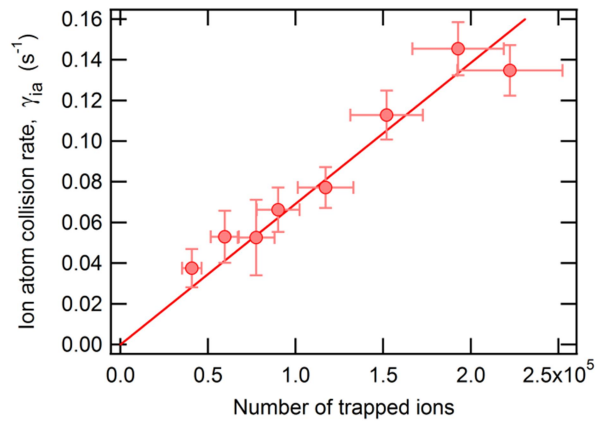

FIG. 6. The atom-ion collision rates $\gamma_{ia}$ plotted against the number of trapped ions $N_i$. The linear fit (solid line) shows that the value of $\gamma_{ia}$ increases linearly with $N_i$.

We use SIMION® to simulate the trajectory of an ion in the ion trap. Collisions with background atoms, MOT atoms and other ions are entirely neglected. The largest time step in the simulation is 0.01 μs, a factor of 233 smaller compared to the rf time period 2.33 μs. In order to determine $V_i$, we need to determine the radial extent ($r_i$) and axial extent ($z_i$) of the ions. We determine $r_i$ by running ion trajectory simulation and recording the lifetime of the ion in the trapping region (i.e. the space circumscribed by the four rods and the two end caps) for different initial positions of the ion. In the simulation, the ion's initial kinetic energy (KE) is $10^{-7}$ eV ($\approx k_B \cdot 1$mK) with the initial velocity vector pointing in the (1,1,1) direction (i.e. making 45° with all three axes) and the initial position is radially displaced in the (1,0,0) direction. The axes orientations are shown in Fig. 1(a). We find that for radial displacements $\geq 2.4$ mm the ion spends less than 1 ms in the trapping region. If we consider an ion to be trapped only if the lifetime exceeds 100 ms, then the estimated value of the radial extend of the ion cloud is $r_i \approx 2.1(\pm 0.3)$, where the error bar reflects the uncertainty in the choice of the cut-off lifetime. The simulations with initial displacements in the (0,1,0) and (1,1,0) resulted in $r_i$ values consistent with the one quoted above.

We determine the ion trap depth $D \approx 0.8(\pm 0.2)$ eV from the expression $D = m_i \omega_r^2 r_i^2/2$, where $\omega_r$ is the radial secular frequency and $m_i$ is the mass of the ion. The radial secular frequency $\omega_r \approx 2\pi \cdot 80$ kHz and the axial secular frequency $\omega_z \approx 2\pi \cdot 23$ kHz are determined from the periodicities of the ion trajectory simulated in SIMION®. With the value of $D$ as input, we determine the axial extent of the ions $z_i \approx 7.3(\pm 1.0)$ mm from the expression $z_i = \sqrt{2D/m_i\omega_z^2}$. Here, we have assumed that energy is equally partitioned among the secular modes of the ion because in the experiment the ion undergoes collisions that keep its velocity randomized. If we now assume the ion cloud shape to be ellipsoidal, we get $V_i \approx 1.35(\pm 0.5) \times 10^{-7}$ m$^{-3}$.

So far, we have considered the trajectory of a single ion in the simulations, neglecting the effects of ion-ion repulsion that might be present in the experiment where multiple ions are trapped. The ion-ion interaction is indeed negligible at the low ion density (~$10^{12}$ m$^{-3}$) in our trap since the ion-ion distance is ~100 μm resulting in a very small interaction energy (~15 μeV) compared to the depth of the ion trap (~0.8 eV). Nevertheless, in order to verify that the estimates of $r_i$, $z_i$ and $V_i$ are correct, we perform another SIMION simulation of 100 ions including ion-ion interaction. The ions initially have a 3-dimensional Gaussian distribution within an ellipsoid with $3\sigma$ radii of (2.1, 2.1, 7.3) mm and each ion is given a charge of $+2500\,e$ so that the total charge of the 100 ions is equivalent to the total charge of the largest ion cloud ($N_i \sim 2.5 \times 10^5$) in the experiment. The ions are allowed to evolve and we find that > 90% remain trapped for 50 ms (~ 4000 secular periods). This suggests that there is no appreciable reduction in ion lifetime due to the presence of ion-ion interaction. This generates confidence in the value of $r_i$, $z_i$ and $V_i$ determined using a single ion.

We determine $k_{ia} = 9.3(\pm 0.4)(\pm 1.2)(\pm 3.5) \times 10^{-14}$ m$^3$s$^{-1}$, using the values of $\gamma_{ia}/N_i$ and $V_i$ determined above. The values in the first, second and third parentheses denote the uncertainties arising from statistical error in $\gamma_{ia}/N_i$, systematic error in $N_i$ and systematic error in $V_i$, respectively.

### E. Comparison with calculations

The collision rate coefficient can be calculated from the expression $k_{ia} = \langle \sigma v \rangle$. Here $\sigma$ is the total cross section. For a homonuclear ion-atom system like the one here, this cross section includes elastic, resonant charge exchange and other energetically allowed inelastic processes, such as hyperfine state change. In the energy regime of the present experiment, the cross section is overwhelmingly dominated by the elastic channel ($\sigma^e$), which is then used to estimate the theoretical collision rate coefficient using:

$$\sigma \approx \sigma^e = \pi(m_i C_4^2/2\hbar^2)^{1/3}(1+\pi^2/16)(m_i v^2/2)^{-1/3},$$

where the last equality is taken from Ref. [22] and the value of $C_4 = 1.37 \times 10^{-56}$ Jm$^4$ is calculated from the Cs ground state polarizability reported in Ref. [38]. The angular bracket denotes a weighted average over all relative speeds $v$ accessed in the experiment. The relative ion-atom speed $v$ is taken to be the speed of the ion (since the MOT atoms are essentially at rest) and is calculated assuming a Maxwell Boltzmann (MB) distribution $f(v)$ of ions speeds within the ion trap. The depth of the ion trap, $D \approx 0.8$ eV, implies that $v$ has an upper bound $v_{max} \approx 1070$ m/s, calculated from the expression $m_i v_{max}^2/2 = D$. Therefore, ideally the MB distribution should be such that all (> 99.9%) ions have speeds $< v_{max}$. The MB distribution $f(v)$ that satisfies this condition has a temperature $T_i \approx$ 1100 K and most probable speed $v_{mp} \approx 370$ m/s, which corresponds to a most probable ion energy $E \approx 95$ meV. Using this MB distribution $f(v)$, we calculate $\langle \sigma v \rangle_{6s} = \int \sigma_{6s} v f(v) dv = 9.5 \times 10^{-14}$ m$^3$s$^{-1}$, assuming all Cs atoms to be in the ground $6S_{1/2}$ state. However, in the experiment a fraction $f_e = 0.092(12)$ of the MOT atoms are in the $6P_{3/2}$ state with polarizability, and hence $C_4$, 4.11 times higher than the $6S_{1/2}$ state [38]. On taking this into account, we get $\langle \sigma v \rangle_{MOT} = 10.9(2) \times 10^{-14}$ m$^3$s$^{-1}$. This value is in fair agreement with the experimentally measured value of $k_{ia}$ reported in section III.D.

In the estimation of collision energy we have ignored the sympathetic cooling of Cs$^+$ ions by the Cs MOT atoms [18] since the number of ions in this article is large and the cooling capacity of the MOT atoms is shared among all these ions, leading to negligible cooling per ion. We note that $\langle \sigma v \rangle$ has a weak dependence on collision energy, with $\langle \sigma v \rangle_{MOT}$ varying from $8.8 \times 10^{-14}$ m$^3$s$^{-1}$ at $E = k_B \cdot 300$ K to $12.0 \times 10^{-14}$ m$^3$s$^{-1}$ at $E = k_B \cdot 2000$ K. While it might be tempting to attribute any discrepancy between theory and experiment to incorrect determination of temperature, this should be resisted since, as we have already seen, the largest source of error in the experiment arises from the estimate of $V_i$. It is therefore a worthwhile future goal to invest in finding a better method to determine $V_i$ in order to increase the precision of the $k_{ia}$ measurement. The motivation for investing in the improvement is that it has the potential to refine into a very precise method for the determination of $k_{ia}$, and hence cross section, over a large range of energy.

## IV. CONCLUSION AND DISCUSSION

We report the measurement of low energy (~ 95 meV) collisions between Cs$^+$ ions and Cs atoms using a hybrid trap apparatus. The measurement relies on the detection of fluorescence from the Cs MOT atoms and recording the loading of the Cs MOT in the absence and presence of trapped Cs$^+$ ions. The loading dynamics is described using a rate equation model and excellent fit to the experimental data is observed. The atom-ion collision rate $\gamma_{ia}$ is obtained from the fitted values and it is then used to determine the collision rate coefficient $k_{ia}$, which agrees fairly well with theoretical calculations. As a supplementary outcome, we determine the photo-ionization cross section $\sigma_p$ of Cs atoms in the $6P_{3/2}$ state for 473 nm laser radiation.

In future it will be interesting to perform experiments with heteronuclear atom-ion systems such as Rb-Cs$^+$ and compare the saturated value of $\gamma_{ia}$ with that of Cs-Cs$^+$. The ratio of $\gamma_{ia}$ in the two cases will provide a measure for the relative values of $k_{ia}$, bypassing the need to calculate $V_i$ explicitly. Direct comparison with the theoretically calculated ratio of $k_{ia}$ can be made very reliably. For such an experiment two MOTs, a Rb MOT and a Cs MOT, would need to be operated simultaneously and therefore one has to carefully consider the atom loss due to collisions between the ultracold Rb and Cs atoms. In addition, the ionization light should have wavelength greater than 479 nm (to avoid ionization of Rb atom) but less than 508.2 nm (to ionize the Cs atoms).

## ACKNOWLEDGMENTS

SAR thanks E. Krishnakumar for useful discussions and acknowledges support from the MeitY Center for Excellence in Quantum Technology grant. Support from Department of Science and Technology (DST), India is gratefully acknowledged. SD acknowledges support from the DST-INSPIRE Faculty Award (IFA14-PH-114) and the Department of Atomic Energy, Government of India, under project no. 12-R&D-TFR-5.02-0200.

———————————